\documentclass{emulateapj}

\newcommand{\ts}{\thinspace}
\newcommand{\simless}{\mathbin{\lower 3pt\hbox
     {$\rlap{\raise 5pt\hbox{$\char'074$}}\mathchar"7218$}}}
\newcommand{\simgreat}{\mathbin{\lower 3pt\hbox
     {$\rlap{\raise 5pt\hbox{$\char'076$}}\mathchar"7218$}}}

\newcommand{\about}    {$\sim$\ts}
\newcommand{\aboutless}{$\simless$\ts}

\newcommand{\msun}{\,M$_\odot$}
\newcommand{\mearth}{\,M$_\oplus$}
\renewcommand{\micron}{\,$\mu$m}
\newcommand{\etal}{et~al.}

\newcommand{\coj}{$^{12}$CO J=$3-2$}
\begin{document}

\title{An ALMA Continuum Survey of Circumstellar Disks in the Upper Scorpius OB Association}

\author{John M. Carpenter}
\author{Luca Ricci}
\author{Andrea Isella}

\affil{California Institute of Technology, Department of Astronomy, MC 249-17, Pasadena, CA 91125}

\begin{abstract}

We present ALMA 880\micron\ continuum observations of 20 K and M-type stars in
the Upper Scorpius OB association that are surrounded by protoplanetary disks.
These data are used to measure the dust content in disks around low mass stars
(0.1-1.6\msun) at a stellar age of 5-11\,Myr. Thirteen sources were detected
in the 880\micron\ dust continuum at $\ge3\sigma$ with inferred dust masses
between 0.3 and 52\mearth. The dust masses tend to be higher around the more
massive stars, but the significance is marginal in that the probability of no
correlation is $p\approx0.03$. The evolution in the dust content in disks was
assessed by comparing the Upper Sco observations with published continuum 
measurements of disks around \about 1-2\,Myr stars in the Class~II stage in 
the Taurus molecular cloud. While the dust masses in the Upper Sco 
disks are on average lower than in Taurus, any difference in the
dust mass distributions is significant at less than $3\sigma$. For stellar
masses between 0.49\msun\ and 1.6\msun, the mean dust mass in disks is lower
in Upper Sco relative to Taurus by $\Delta\mathrm{log\, M_{dust}} = 
0.44\pm0.26$.
\end{abstract}


\section{Introduction}

The lifetime of optically thick, gas-rich disks surrounding young 
stars provides empirical constraints on the timescales to form planetary
systems and the mechanisms responsible for disk dispersal.
The disk dissipation timescale is typically measured by surveying clusters or
association of stars of various ages and identifying the fraction of stars that
exhibit infrared emission in excess of the stellar photosphere, which is
attributed to a circumstellar disk that absorbs and re-radiates the stellar
radiation. Infrared surveys have shown that \about 80\% of K- and M-type stars
are surrounded by a disk at an age of \about 1\,Myr, and declines to \aboutless
20\% at an age of \about 5\,Myr \citep{Haisch01,Mamajek04,Hernandez08}. Disks
around A and B-type stars (\about 2-3\msun) appear to evolve on even shorter
timescales \citep{Hernandez05,Carpenter06,Dahm09}.

Submillimeter continuum observations provide additional key diagnostics of disk
evolution. Whereas infrared emission is generally optically thick and
traces the disk surface layer within \about 1\,AU of the star,
submillimeter continuum emission is optically thin over most of the disk and
can also probe the cool, outer disk. The submillimeter continuum emission 
is a measure of the surface area of millimeter-sized particles in the disk 
\citep[e.g.,][]{Ricci10b}, and can be used to estimate the dust mass with 
assumptions on the dust opacity and temperature structure of the disk.

Hundreds of $\sim$ 1-2\,Myr old stars in the Taurus and Ophiuchus clouds have
been surveyed in the submillimeter continuum with single dish telescopes and
interferometers \citep{Beckwith90, Andre94, Motte98, Andrews05, Andrews07b, Schaefer09,
Andrews13}, and the continuum and/or spectral-line emission have been resolved
in dozens of stars \citep{Dutrey96, Simon00, Kitamura02, Andrews07a, Isella09,
Andrews09, Andrews10, Kwon11, Guilloteau11}. Collectively, these extensive
observations have established the disk properties around low-mass stars at an
age of \about 1-2\,Myr.

Submillimeter observations of stars at other ages are more limited, but
nonetheless have begun to reveal how the dust mass evolves. Submillimeter and
millimeter observations of the \about 2-3\,Myr IC~348 \citep{Carpenter02,Lee11}
and the \about 5-11\,Myr Upper Scorpius OB association \citep{Mathews12a}
demonstrate that these regions lack the luminous disks found in Taurus and
Ophiuchus. However, the stellar samples observed so far in IC~348 and Upper Sco
are incomplete, and \citet{Andrews13} have suggested that the lack of bright
disks may be due to a selection bias toward late type stars rather than to 
intrinsically different distribution of disk submillimeter luminosities. 
After considering the lower mean submillimeter flux density observed
in disks around lower
mass stars, \citet{Andrews13} showed that the millimeter-wavelength luminosity
distribution of the IC 348 and Taurus samples are statistically
indistinguishable, while the Upper Scorpius OB Association (hereafter Upper
Sco) sample appears to have only marginally ($\sim 2.5\sigma$) lower
luminosities on average. 

More recently, \citet{Williams13} presented a large submillimeter survey of
disks in the \about 3\,Myr old $\sigma$ Orionis cluster. In this case they found
that the submillimeter luminosities are lower in $\sigma$ Orionis than in
Taurus, indicating a decline of the amount of material in disks as star-forming
regions age from $\sim 1$ to $\sim 3$\,Myr.

We report new submillimeter continuum observations of K and M-type stars in 
Upper Sco obtained with ALMA during Cycle 0 Early Science. These
data achieve nearly an order of magnitude better sensitivity than previous
submillimeter surveys of disks in Upper Sco. We use these data to investigate
any dependence of the disk properties with stellar mass, and compare these
observations with existing submillimeter continuum measurements of stars in the
younger Taurus region to investigate the evolution of dust masses.

\section{The Upper Sco Sample}
\label{sample}

The initial sample consisted of 24 K and M-type stars in Upper Sco that were 
identified with
an infrared excess between 3.6\micron\ and 16\micron\ by \citet{Carpenter06}.
The characteristics of the infrared excess suggest that these stars are
surrounded by optically thick disks in the Class II evolutionary stage
\citep{Lada84}. Table~\ref{tbl:sample} lists the 20 sources observed with 
ALMA, the phase center of the ALMA observations, and the date of the ALMA
observations. The four sources that were not
observed before the end of Cycle 0 are 
%
J161115.3$-$175721 (M1 spectral type),
%
J160545.4$-$202308 (M2),
%
J160357.9$-$194210 (M2),
and 
J160953.6$-$175446 (M3).

The stellar luminosity ($L_*$), effective temperature ($T_*$), and mass ($M_*$)
were estimated based on available photometry and spectroscopy. The effective
temperature scale was set based on the observed spectral type. In anticipation
of comparing the ALMA observations of Upper Sco with observations of Taurus
presented in the literature, the temperature scale described in
\citet{Andrews13} and references therein was adopted. 

Observed optical and near-infrared photometry were drawn from the 2MASS
\citep{Skrutskie06,Cutri03} and DENIS \citep{DENIS} photometric catalogs. The visual extinction was
estimated from the observed DENIS $I-J$ color by adopting the intrinsic colors
for 5-30\,Myr stars from \citet{Pecaut13} and the \citet{Cardelli89} extinction
law. DENIS photometry was not available for 3 sources, for which we adopt
$A_V=0.7$\,mag (the median value for the remaining stars) and an uncertainty of
0.5\,mag. For J161420.2-190648, the visual extinction derived from the DENIS
$I-J$ color ($\mathrm{A_V}=4.0\pm0.27$\,mag) produced a visual extinction
significantly larger than derived by \citet{Preibisch02} from $R-I$ photometry
($\mathrm{A_V}=1.8$). This star exhibits excess emission in the near-infrared
bands that may contribute to the $J$-band photometry \citep{Dahm09}; we adopted
$\mathrm{A_V}=2\pm0.5$\,mag for this star.

The uncertainties in the effective temperatures assume a spectral type
uncertainty of $\pm$1 subclass. The uncertainties in the luminosities include
the uncertainty in the $J$-band photometry, the extinction, and the distance to
Upper Sco, which is assumed to be 15\% of the mean distance of 145\,pc
\citep{deZeeuw99}. Assuming gaussian distributions in log~$L_*$ and log~$T_*$,
the distribution of possible stellar masses (and ages) were derived using the
\citet{Siess00} pre-main-sequence evolutionary tracks with a metallicity of
Z=0.02 and no convective overshoot. The stellar mass distribution was then
inferred by marginalizing over the stellar ages. Table~\ref{tbl:stellar} lists
the derived stellar parameters.

\section{ALMA Observations}

The ALMA Early Science Cycle 0 observations were obtained on 2012 Aug 24 UT (7
sources), 2012 Aug 28 (6 sources), and 2012 Dec 16 (5 sources).
Table~\ref{tbl:obs} summarizes the observations, including the number of 12\,m
antennas used, the minimum and maximum project baselines, the primary flux
calibrator, a secondary flux calibrator, the passband calibrator, and the gain
calibrator for each day. 

All observations were obtained in Band 7 with a full-width-at-half-maximum
(FWHM) primary beam size of 18.5\arcsec. The correlator was configured to
record dual polarization for spectral windows centered on 333.8, 335.7, 345.8,
and 347.7\,GHz for a mean frequency of 340.7\,GHz (880\micron). Each window
provided a bandwidth of 1.875\,GHz with channel widths of 0.488\,MHz. The
spectral resolution is twice the channel width. One spectral window was
centered on the \coj\ line at a rest frequency of 345.79599\,GHz. The channels
with \coj\ emission were omitted when analyzing the continuum data. 

The ALMA data were calibrated using the CASA package. The initial reduction
scripts were kindly provided by Crystal Brogan (NRAO), which included 
phase calibration with the 183\,GHz water vapor radiometers, bandpass 
calibration, flux calibration, and gain calibration. Table~\ref{tbl:obs} 
lists the calibrators for each night. We adapted the initial calibration
scripts to perform bandpass, flux, and gain calibration using CASA 4.1. 

Flux calibration was established by observing either Neptune or Titan, and
adopting the Butler-JPL-Horizon 2012 models. Due to the broad \coj\ absorption 
line present in the atmospheres of Neptune and Titan at 345.8\,GHz, we measured 
the flux densities of the bandpass and gain calibrators only in the 333.8 GHz
and 335.7\,GHz spectral windows. The flux densities measured in these two
windows were consistent to within 3\% for a given source on a single day. The
average flux density in these two windows was adopted for all 4 spectral
windows.

The measured flux densities of the passband, secondary, and gain calibrators
were 16\% brighter on average for the 2012 Aug 28 data than on 2012 Aug 24.
Given the measurements were obtained 4 days apart and the flux differences were
common to 3 different calibrators, we assume that this represents a systematic
difference in the absolute flux calibration between the two datasets.
For these two days, we averaged the two flux measurements for each
calibrator. The adopted flux density for the gain calibrator J1625$-$2527 on
these two nights was 0.97\,Jy. We adopt a 1$\sigma$ calibration uncertainty of
10\%.

Images were created from the calibrated visibilities using CASA 4.1 using a
Briggs robust weighting parameter of 2. A continuum map was produced by
averaging all of the channels except those around the \coj\ line. The 1$\sigma$
point source sensitivity near the phase center
is typically 0.19, 0.16, and 0.52\,mJy~beam$^{-1}$ for sources
observed on 2012 Aug 24, 2012 Aug 28, and 2012 Dec 16, respectively.

\section{ALMA Results}
\label{flux}

\begin{figure}
\plotone{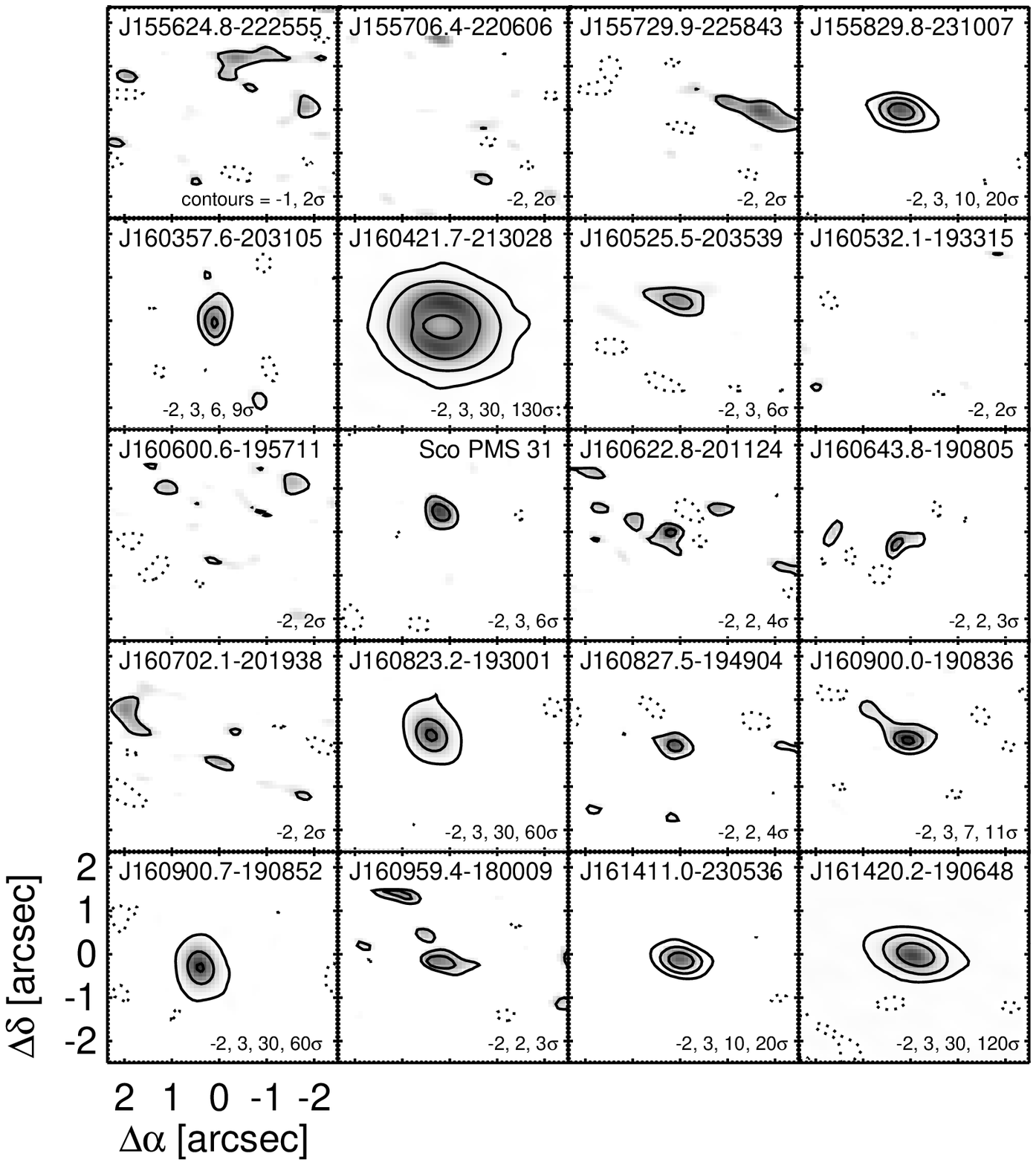}
\caption{
Contour maps of the ALMA 880\micron\ continuum emission for 20 K-M type
stars in Upper Sco. Each map is centered on the stellar position after 
correction for proper motion. The contour levels are indicated in the 
lower right for each panel, where solid and dotted contours indicate 
positive and negative flux densities, respectively.
}
\label{fig:images}
\end{figure}

Figure~\ref{fig:images} presents contour maps on the 880\micron\ 
continuum emission for the Upper Sco sample. Each image is centered on the
expected stellar position, which was
computed using the coordinates and proper motions in the PPMXL \citep{Roeser10}
catalog. The median offset of the expected stellar position from the phase
center of the ALMA observations is $(\Delta\alpha, \Delta\delta) = (-0.17'',
-0.33'')$. 

\begin{figure}
\plotone{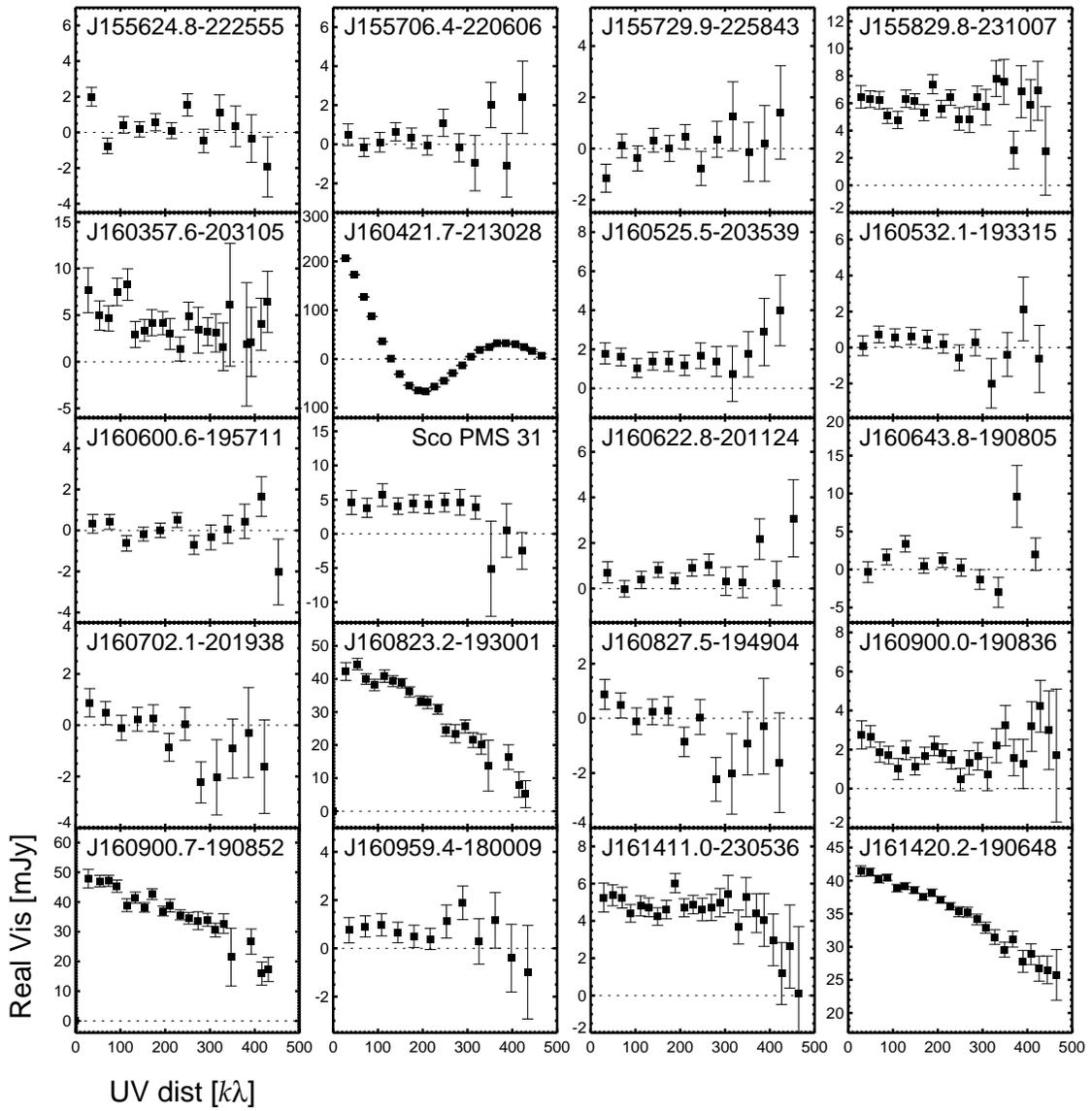}
\caption{
The real part of the visibility as a function of projected baseline length
for the ALMA 880\micron\ continuum data for 20 K-M type stars in Upper Sco. 
The phase center has been shifted to correspond to the centroid of the 
continuum emission, or the stellar position if the continuum is not detected.
}
\label{fig:uvreal}
\end{figure}

Figure~\ref{fig:uvreal} presents the real component of the observed continuum
visibilities for each source. The continuum emission toward 4 sources
(J160421.7-213028, J160823.2-193001, J160900.7-190852, and J161420.2-190648)
are clearly resolved in that the visibilities decline in amplitude with
increasing $uv$ distance: The 4 resolved sources are also the brightest disks
in the sample with flux densities in excess of 40\,mJy at 880\micron; the
remaining sources have flux densities less than 6\,mJy. The dust emission
around J160421.7-213028 is resolved into a ring, which was previously imaged
in the  850\micron\ continuum with the SMA \citep{Mathews12a} and scattered 
light \citep{Mayama12}. Zhang \etal~(2014, in preparation) present an extensive
analysis of the ALMA continuum and molecular line data for this source. 
Model fitting for the remaining sources and along with analysis of the
\coj\ data will be presented in a separate paper.

Flux densities were measured by fitting an elliptical gaussian to the
visibility data using {\em uvmodelfit} in CASA. The model contains 6 free
parameters: 1) the integrated flux density, 2) the full-width-at-half-maximum
(FWHM), 3) the aspect ratio, 4) the position angle, 5) the right ascension
offset from the phase center, and 6) the declination offset from the phase
center. The uncertainty on the model parameters were scaled by the factor 
needed to produce a 
reduced chi-squared of unity. If the ratio of the FWHM to the uncertainty in
the FWHM was less than 2, a point source model with three free parameters
(integrated intensity and position offsets) was fitted to the visibility data
instead. For the disk ring around J160421.7$-$213028, the flux density was
measured using aperture photometry in a circular aperture of radius 1.5\arcsec\
in the deconvolved image.

Table~\ref{tbl:continuum} summarizes the continuum measurements.
The table includes the integrated flux density, the offset of the submillimeter
emission from the stellar position, the rms noise in the synthesized image, and
the FWHM and position angle of the deconvolved beam. The uncertainty in the
offsets include the uncertainties in the stellar position at the measured epoch
in the PPMXL catalog, the proper motion propagated since that epoch, and the
centroid of the submillimeter continuum emission. Upper limits to the flux
density were computed as $\mathrm{max}(0, S_\nu) + 3\times$rms, assuming that
the emission originates from a point source.

Thirteen sources were detected in the 880\micron\ continuum with a
signal-to-noise ratio of 3 or greater. Within the 3$\sigma$ uncertainties, the
centroid of the continuum emission is consistent with the stellar position,
with a median offset between the centroid of the 880\micron\ continuum and the
expected stellar position of ($\Delta\alpha, \Delta\delta$) = ($0.02'',
0.05''$). We conclude that most of the ALMA detections must be associated with
the star, and that there are no clear examples of extragalactic contamination
in the sample. We assume throughout this paper that the detected continuum
sources are associated with the Upper Sco stars.

\section{Properties of Disks in Upper Sco} 

In this section, the ALMA continuum measurements are used to infer the mass of
dust in the circumstellar disks. After describing how the dust masses are
estimated, we examine if the dust masses vary systematically with stellar mass
within Upper Sco. We then compare the dust mass distribution with stars in the
younger Taurus molecular cloud to constrain the evolution of dust mass with
time.

\subsection{Dust Masses}

Interferometric observations can be used to measure the dust masses in disks
by fitting a parameterized surface density model to the observed visibilities.
While interferometric data are available for the entire Upper Sco sample, such
data are not available for all sources in the Taurus comparison sample
described below. Therefore, we adopt a simplified approach to estimating dust
masses that can be applied to all sources, as outlined in \citet{Andrews13}.
Assuming the dust emission is isothermal and optically thin,
the dust mass is given by
\begin{equation}
\mathrm{log}\,M_\mathrm{dust} = \mathrm{log}\,S_\nu + 2\,\mathrm{log}\,d - 
\mathrm{log}\,\kappa_\nu - \mathrm{log}\,B_\nu(T_\mathrm{dust}),
\end{equation}
where $S_\nu$ is the observed flux density, $d$ is the distance, $\kappa_\nu$
is the dust opacity, and $B_\nu(T_\mathrm{dust})$ is the Planck function for
the dust temperature $T_\mathrm{dust}$. We adopt $d=145$\,pc, which is the mean
distance of the OB stars in the Upper Sco association \citep{deZeeuw99}. For
consistency with \citet{Andrews13}, we adopt $\kappa_\nu$=2.3\,cm$^2$~g$^{-1}$
at 230\,GHz, and assume $\kappa_\nu$ scales with frequency as $\nu^\beta$,
where $\beta=0.4$. The dust temperature is estimated as
$T_\mathrm{dust} = 25\,\mathrm{K} \times(L_*/L_\odot)^{0.25}$
\citep[see][]{Andrews13}. While a range of dust temperatures will be present 
in a disk, this formalism represents the characteristic dust temperature
that describes the continuum emission.

Table~\ref{tbl:mdust} lists the derived dust masses. The uncertainties in the
dust mass include the uncertainties in the measured flux densities and the
distance to Upper Sco. The dust mass uncertainties do not include errors in the
assumed dust opacity. However, the relative changes in the inferred dust masses
may be more accurate to the extent that the dust properties are similar between
disks.

\subsection{Dust Mass vs. Stellar Mass in Upper Sco}
\label{mdustmstar}

Figure~\ref{fig:mdust_usco} shows the derived dust masses as a function of the
stellar mass for the 20 stars in Upper Sco. The inferred dust masses of the
sources detected with ALMA range over two orders of magnitude from
0.3\mearth\ to 52\mearth, which represents \about 0.01\% to 1.7\% of the
stellar mass assuming a dust-to-gas ratio of 0.01 by mass. Considering both 
detections and upper limits, most disks have dust masses less than 1\mearth.

Disks around lower mass stars tend to have lower
dust masses than the disks around higher mass stars. Eight of the 9 stars with
spectral type M3 or earlier (M$_* > 0.26$\msun) were detected with ALMA. The
one star not detected was one of the five stars that had lower sensitivity
compared to the majority of the sample. By comparison, of the eleven M4 and M5
stars in the sample, only 5 were detected, even though all 11 stars had
high-sensitivity data. Thus the predominant number of stars with non-detections
are the late spectral types with lower stellar masses. 

The significance of these apparent trends were evaluated using the correlation
tests adapted for censored data sets \citep{Isobe86} as implemented in the
ASURV software package \citep{Lavalley92}. The Cox proportional hazard test,
the Kendall rank test, and the Spearman rank test in ASURV
indicate that the probability of no correlation between dust mass and stellar
mass is 0.017, 0.044, and 0.039 respectively. We conclude that there is
marginal evidence that the dust masses declines with stellar mass in the Upper
Sco sample. 

Assuming a power-law relation is present between the dust masses and the
stellar masses, the slope of the power-law was derived using the Bayesian
method described in \citet{Kelly07} that takes into account the measurement
uncertainties, upper limits, and intrinsic scatter in the relationship. The
derived relationship between the dust mass and the stellar mass is given by
$\mathrm{log} (M_\mathrm{dust}/\mathrm{M_\oplus}) = (0.68\pm0.30) +
(1.01\pm0.60)\, \mathrm{log} (M_*/\mathrm{M_\odot})$ with a spread of
$0.76\pm0.24$\,dex. Thus the slope is consistent with a linear correlation
between dust mass and stellar mass, but the uncertainties on the slope are such
that no correlation is consistent with the data. The slope and spread for Upper
Sco are consistent with that for disks in Taurus (slope=$1.1\pm0.4$, dispersion
= $0.7\pm0.1$\,dex) found by \citet{Andrews13}, which is shown as the shaded
region in Figure~\ref{fig:mdust_usco}. In the following section, we present a
more quantitative comparison between the Upper Sco and Taurus samples.

\section{Comparison between Upper Sco and Taurus}
\label{comparison}

The percentage of K- and M-type stars in Upper Sco that retain an optically
thick inner disk is \about 19\% \citep{Carpenter06} and may vary with stellar
mass within this spectral type range \citep{Luhman12}. Given that \about 
80\% of low-mass stars with an age of \about 1\,Myr contain a disk
\citep{Hernandez08}, the average disk mass must be lower in Upper Sco compared
to younger regions. However, the question remains if the dust masses stay
relatively constant before dispersing rapidly, or if there is a steady decline
in the dust mass as it disperses. These different scenarios ultimately reflect
the mass loss rate in the disk and the mechanisms responsible for the disk
dispersal. We aim to quantify this evolution by comparing submillimeter
continuum observations towards stars of various ages that still retain
optically thick disks.

In addition to the Upper Sco observations presented here (see also
\citealt{Mathews12a}), other star forming regions that have been surveyed at
submillimeter wavelengths include Taurus \citep{Beckwith90, Andrews05,
Andrews13}, $\rho$~Oph \citep{Andre94, Motte98, Andrews07b}, IC~348
\citep{Carpenter02, Lee11}, the Orion Nebula Cluster
\citep{Mann09a,Mann09b,Mann10,Eisner08} NGC~2024 \citep{Eisner03}, MBM~12
\citep{Itoh03, Hogerheijde02}, Lupus \citep{Nuernberger97}, Chamaeleon~I
\citep{Henning93}, Serpens \citep{Testi98}, and $\sigma$ Orionis
\citep{Williams13}. Taurus is the one region that can be most readily compared
with Upper Sco for a number of reasons. After decades of searching for members
(see recent compilations by \citealt{Rebull10} and \citealt{Luhman10}),
the stellar census is likely nearly complete for sources with and without
disks. A wealth of ancillary data, including spectral types, are available for
most members, so that a robust comparison can be made with Upper Sco over the
same stellar mass range. Finally, the close proximity leads to improved
sensitivity, as most disks around stars in Taurus that have spectral types
earlier than M3 have been detected in the submillimeter continuum
\citep{Andrews13}.

\subsection{Relative ages}

The age of Upper Sco is commonly assumed to be \about 5\,Myr based on the
kinematic of the B-type stars \citep{Blaauw78}, and placing association members
in an HR diagram and inferring the age from evolutionary tracks
\citep{deGeus89,Preibisch02,Slesnick08}. More recently \citet{Pecaut12} derived
an age of 11$\pm2$\,Myr for Upper Sco based on the isochronal ages of the B, A,
and G-type stars and the M supergiant Antares, and the luminosities of the
F-type stars. They also derive a lower limit of 10.5\,Myr (99\% confidence) on
the expansion age using radial velocities and {\em Hipparcos} parallaxes.

By comparison, the mean age of stars with disks in Taurus is \about 1-2\,Myr,
as inferred by placing stars in an HR diagram
\citep[e.g.,][]{Kenyon95,Hartmann01,Bertout07,Andrews13}. Thus Upper Sco
appears older than Taurus. While one must be cautious of ages derived by
different techniques, qualitative signatures also support the notion that Upper Sco
is older. First, the natal molecular cloud has been dispersed as the visual
extinction toward the association members is typically $\mathrm{A_V}<2$\,mag.
Also, the association lacks stars in the Class~0 and Class~I phases which
typify young star forming regions \citep[e.g.,][]{Gutermuth11}. Finally,
late-type stars in Upper Sco have surface-gravity photospheric spectral lines consistent
with an older age relative to Taurus, and in fact has been used as a defining
characteristic of membership \citep[e.g.,][]{Slesnick06}.
Thus even though the age of Upper Sco may be uncertain by a factor of \about 2,
the association is almost certainly older than Taurus.

\subsection{Relative dust masses}

The comparison sample in Taurus consists of Class II sources compiled by
\citet[see also \citealt{Rebull10}]{Luhman10}. The submillimeter flux densities
for this sample are presented in \citet{Andrews13}, who used new and published
submillimeter observations at multiple wavelengths to estimate the flux density
at a wavelength of 890\micron. The Taurus submillimeter flux densities were
extrapolated to the mean wavelength of the Upper Sco observations (880\micron)
by assuming the dust emission varies with frequency as $\nu^{2.4}$, which is
the same frequency dependence adopted in \citet{Andrews13}. 

It should be noted the upper limits to the submillimeter flux densities for the
Taurus and Upper Sco observations are not computed consistently. Upper limits
in Taurus are generally reported as 3 times the rms noise of the observations,
while the upper limits in Upper Sco derived here are given as 3 times the rms 
plus any positive measured flux density. Thus the upper limits in Upper Sco are
more conservative. Given the expectation that dust masses may be lower in the
Upper Sco due to the older age, the different treatments of the upper limits
will only weaken any differences in the two samples.

We analyzed the samples in two stellar mass ranges: 0.097-0.26\msun\ that
encompasses the M3-M5 stars in Upper Sco, and 0.49-1.6\msun\ that encompasses
the K2-M0.5 stars. The selection was done by stellar mass rather than spectral
type since given the age differences between Upper Sco and Taurus, there is not
a strict correspondence between spectral type and stellar mass. The stellar
masses for the Taurus sample were derived from the effective temperatures and
stellar luminosities in \citet{Andrews13} using the interpolation
procedure adopted for Upper Sco (see Section~\ref{sample}). The difference in
$\mathrm{log}\,M_*$ derived here and those reported in \citet{Andrews13} is
0.00 with a standard deviation of 0.05\,dex. 

The mass ranges of the two samples was motivated by three considerations. First,
the observed Upper Sco sample contains no stars with spectral types between
M0.5 and M3, which leads to a break in the stellar mass distribution. Second,
\citet{Andrews13} found a correlation between dust mass and stellar mass in
that dust masses in disks in Taurus are \about 7$\times$ larger in the
higher mass bin than the lower mass bin. Finally, the fraction of stars with
optically thick disks as traced by infrared observations increases with 
decreasing stellar mass in Upper Sco \citep{Carpenter06,Luhman12} and other
regions (IC~348 - \citealt{Lada06}; NGC~2262 - \citealt{Dahm07}). The
variations in the disk mass and disk lifetime with stellar mass suggests that
the disk mass-loss rate may also vary with stellar mass. 

Ideally any comparison between Upper Sco and Taurus will consider the
multiplicity of the stars since close companions can shorten the lifetime of
disks \citep{Jensen96,Cieza09,Harris12,Kraus12}. However, multiplicity
information is currently available for only 7 of the 20 Upper Sco stars
observed with ALMA \citep{Kraus08}. Therefore, we considered all of the stars
in the Taurus sample that are within the appropriate stellar mass range. This
could potentially bias the results if the remaining disks in Upper Sco are
preferentially found around single stars.

\begin{figure}
\plotone{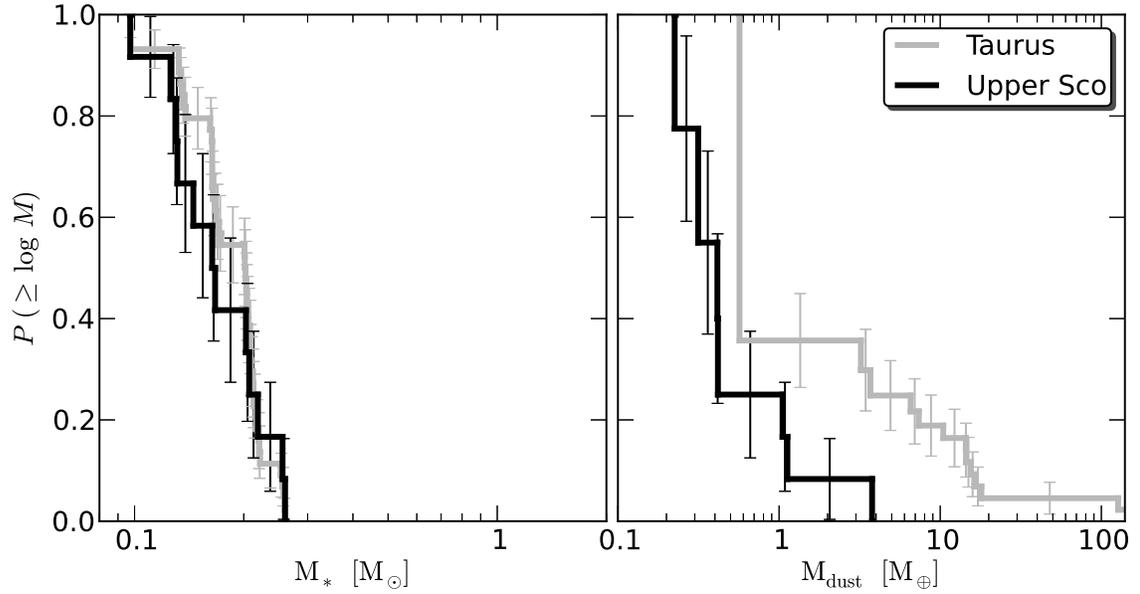}
\caption{
{\em Left:} Cumulative distribution of the stellar masses
for the Taurus (grey) and Upper Sco (black) samples for stellar
masses between 0.097 and 0.26\msun. 
{\em Right:} Cumulative distribution of the dust masses for the
stellar samples shown in the left panel.
Both the stellar masses and dust masses in Upper Sco and Taurus
are consistent with being drawn from the same parent distribution.
}
\label{fig:kmlate}
\end{figure}

\begin{figure}
\plotone{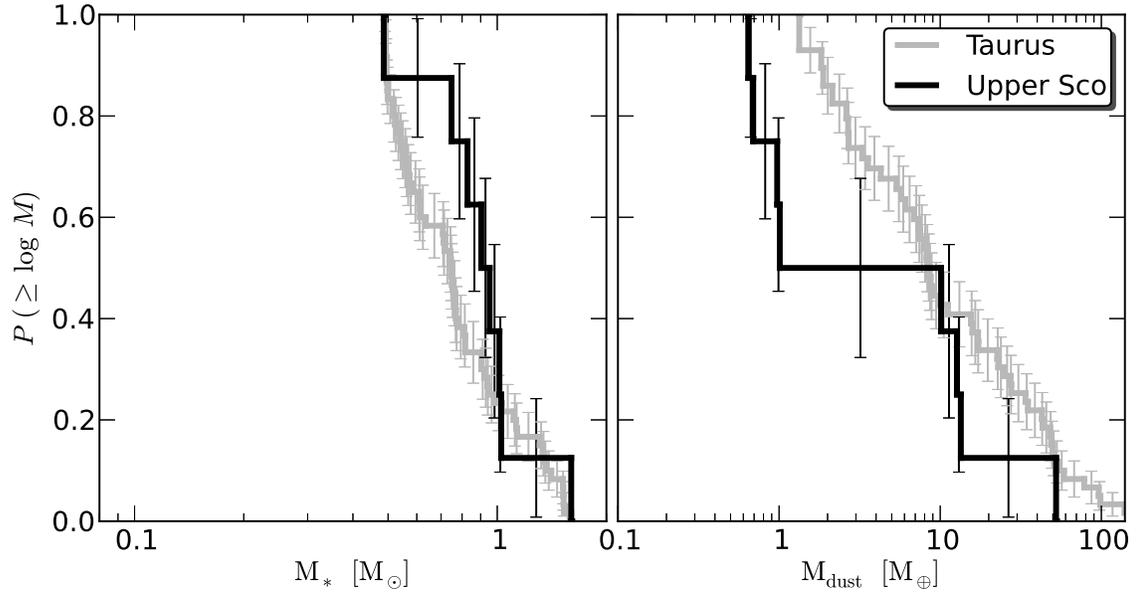}
\caption{
Same as in Figure~\ref{fig:kmlate}, but for stellar masses between
0.49 and 1.6\msun. Both the stellar masses and dust masses in Upper Sco 
and Taurus are consistent with being drawn from the same parent distribution.
}
\label{fig:kmearly}
\end{figure}

We used the two-sample tests in the ASURV package to compare the Upper Sco and
Taurus samples. The left panel in Figure~\ref{fig:kmlate} compares the 
cumulative distribution of stellar masses in Taurus and Upper Sco for the 
lower mass stars. For these stars, the median stellar mass in Taurus is 25\%
higher (0.20\msun\ vs. 0.16\msun) than in Upper Sco. However, the two-sample 
tests in the ASURV package indicate a probability of 0.37-0.96 that the 
distribution of stellar masses are drawn from the same parent population. 
Similarly, for stars in the higher mass bin (see left panel in 
Figure~\ref{fig:kmearly}), the median stellar mass in the Taurus sample is 18\%
lower than in Upper Sco, but the probability that the stellar distributions are
drawn from the same distribution is between 0.20 and 0.26. Thus there is no
evidence for differences in the stellar mass distribution of Class~II sources
Taurus and Upper Sco for the two stellar mass ranges. Therefore, the dust mass
distributions can be reliably compared between the two samples.

The right panel in Figure~\ref{fig:kmearly} shows the cumulative distributions
of the dust masses around the lower mass stars in the Taurus and Upper Sco
samples as estimated using the Kaplan-Meier estimator to factor in upper
limits. Formally the mean dust mass in Upper Sco is $<\mathrm{log
(M_{dust}/M_\oplus)}> = -0.31 \pm 0.10$, which is lower than the mean mass in
Taurus of $<\mathrm{log (M_{dust}/M_\oplus)}> = 0.22 \pm 0.12$. However, only 7
of the 12 sources in Upper Sco were detected with ALMA, and only 12 of the 44
stars in Taurus. The Kaplan-Meier estimator requires that the censored data
points be randomly distributed, which may not be valid for the lower mass
stars. Thus the mean dust masses for the lower mass stars should be treated
with caution. The ASURV two-sample tests provide a robust comparison between
the Upper Sco and Taurus samples that factor in the upper limits from the
continuum observations and does not require that the censorship be random.
These tests indicate the probability that the dust masses in the low mass stars
in Taurus and Upper Sco are drawn from the same parent population is between
0.064 and 0.086. Thus there is only marginal evidence that the lower-mass stars
in the Upper Sco sample have lower disk masses than comparable stars in Taurus.

For the higher stellar mass bin, 7 of the 8 stars in Upper Sco and 48 of
60 stars in Taurus were detected in the submillimeter continuum. Thus the
mean dust masses is more accurately defined
and the median dust mass is robustly determined. The mean dust mass 
is 
$<\mathrm{log (M_{dust}/M_\oplus)}> = 0.57 \pm 0.25$ for Upper Sco and 
$<\mathrm{log (M_{dust}/M_\oplus)}> = 1.01 \pm 0.08$ for Taurus. 
The median $\mathrm{log (M_{dust}/M_\oplus)})$ is higher in Upper Sco compared
to Taurus (0.00 and 0.93, respectively). Thus the dust mass distribution is 
skewed toward lower masses in Upper Sco compared to Taurus. However, the
two-sample tests in ASURV indicate that the probability that the samples are
drawn from the same parent population is between 0.03 and 0.21. Therefore, the
observed differences are not significant.

\section{Conclusions}

The results presented in Section~\ref{comparison} indicate that the
distribution of dust masses between Taurus and Upper Sco are statistically
indistinguishable given the present sample sizes. 
To place limits on the differences in the mean dust mass
between Taurus and Upper Sco, we used the mean dust masses values from the
Kaplan-Meier estimator presented in Section~\ref{comparison}. We consider only
the higher mass stars (0.49-1.6\msun) given the preponderance of upper limits
in Upper Sco and especially Taurus for the lower mass stars. The change in the
mean dust mass from Taurus to Upper Sco for the 0.49-1.6\msun\ stars is
$\Delta\mathrm{log\,M_{dust}} = 0.44\pm0.26$. Thus the mean dust mass has
declined by a factor of $\approx 2.8\pm1.6$, but, consistent with the analysis
presented in Section~\ref{comparison}, the uncertainties are such that no
decline in the mean dust mass is consistent with the data. The 3$\sigma$ upper
limit to the change in the mean $\mathrm{log} M_\mathrm{dust}$ is 1.22\,dex,
and thus formally, these data cannot exclude an order of magnitude change in
the mean dust mass.

\begin{figure}
\plotone{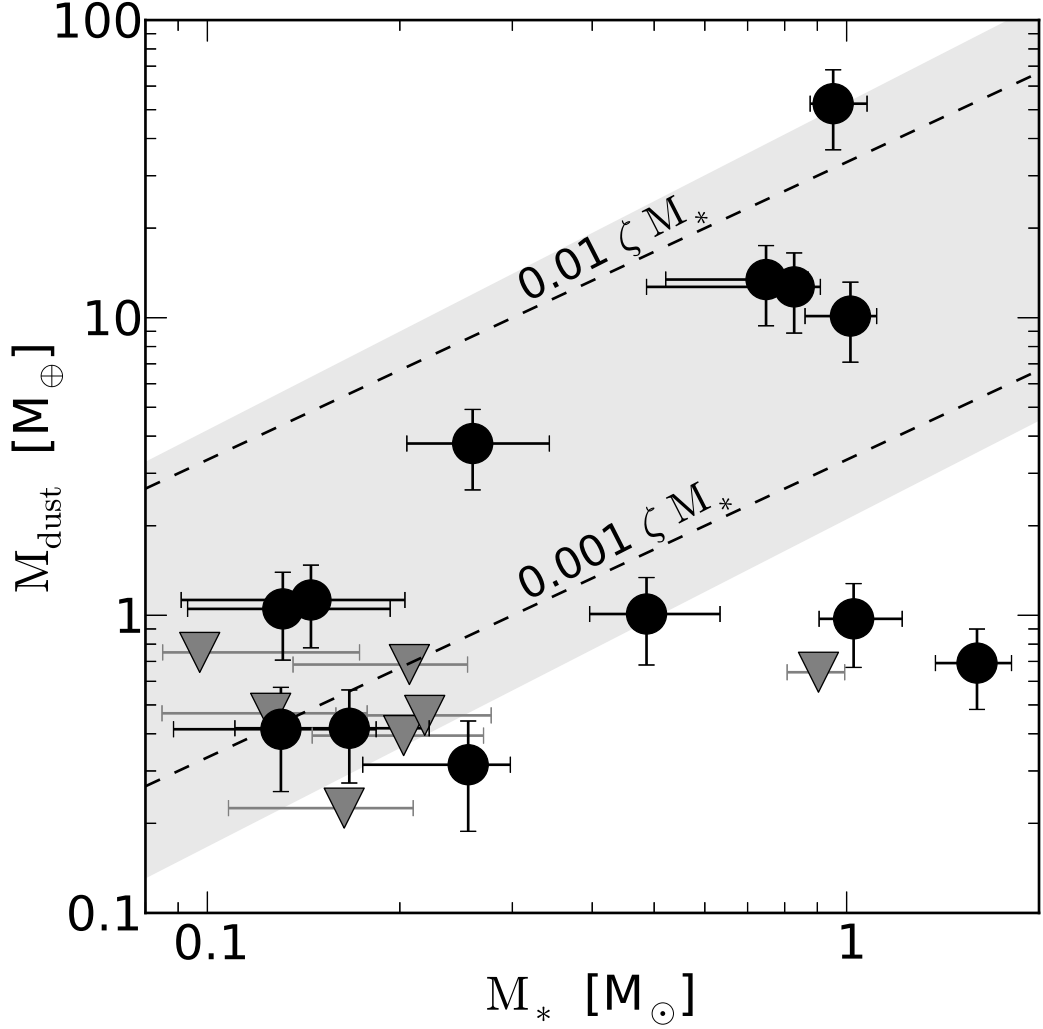}
\caption{
Dust mass as a function of stellar mass for the 20 stars in the Upper Sco
sample. Sources with 880\micron\ continuum detections are denoted by circles,
and 3$\sigma$ upper limits to the 880\micron\ detections are indicated by
triangles. The dashed lines shows constant ratios of dust mass to stellar mass 
of 0.01$\zeta$ and 0.001$\zeta$, where $\zeta=0.01$ is the dust to gas ratio. 
The shaded region shows the correlation between dust mass and disk mass 
derived by \citet{Andrews13} in Taurus, where the width of the region shows 
the 0.7\,dex spread about the best fit relation.
}
\label{fig:mdust_usco}
\end{figure}

The reason why the constraints on the mean dust mass remain poor can be readily
ascertained from Figure~\ref{fig:mdust_usco}. For the 0.49-1.6\msun\ stars,
half have dust masses between \about 10 and 50\mearth\ and half
have masses less than 1\mearth. The gap in the dust
mass distribution within this stellar mass range implies that the median disk
properties remain uncertain by an order of magnitude.

While the lower mean flux densities in Upper Sco relative to Taurus have been
interpreted as a decrease in the dust masses, systematic differences in the
dust composition or the grain size distribution can also lead to a decrease in
the submillimeter flux density for a constant mass in solids. As an example, we
computed the dust opacity by adopting the three most abundant species in the
\citet{Pollack94} dust composition and assuming that the size distribution of
particles can be represented by a power law of $n(a) \propto a^{-3.5}$.
Increasing the maximum particle radius to 1\,cm from 1\,mm, but keeping the
total mass in solids constant, would decrease the observed submillimeter flux
density by a factor 2.7, which is consistent with the observed decrease in the
flux density in Upper Sco relative to Taurus. In this scenario, the slope of
the dust opacity between wavelengths of 1\,mm and 3\,mm will decrease to
$\beta=0.66$ from $\beta=0.91$. While a systematic change of $\beta$ with
stellar age has not been observed \citep{Ricci10a,Ubach12}, the uncertainties
on the measurements for individual disks are typically $\Delta\beta \approx
0.2-0.4$ (1$\sigma$). Thus we cannot exclude the possibility that the size
distribution of particles is changing between Upper Sco and Taurus but the
overall mass of solids has remained the same. Sensitive, long wavelength
observations can help break the degeneracy between variations in grain growth
and disk mass in accounting for the reduced submillimeter flux.

The sample for these observations was drawn from the Spitzer survey presented in
\citet{Carpenter06} for a subset of the known Upper Sco population. Since 
that time, not only has the census of Upper Sco members been refined, 
the all-sky WISE survey between 3.5 and 22\micron\ has been completed, which
can be used to assess the presence of a disk in all association members. 
Such a census has already by completed \citep{Rizzuto12,Luhman12}, and
there are over 200 stars and brown dwarfs over all spectral types in Upper Sco
that have an infrared excess characteristics of a disk, including primordial
and debris disks. Future observations of this large sample with ALMA will
probe the tentative correlations identified in this paper.

\acknowledgements

We thank Crystal Brogan and Steve Myers for their assistance in the data
reduction. The National Radio Astronomy Observatory is a facility of the
National Science Foundation operated under cooperative agreement by Associated
Universities, Inc. This paper makes use of the following ALMA data:
ADS/JAO.ALMA\#2011.0.00966.SSB. ALMA is a partnership of ESO (representing its
member states), NSF (USA) and NINS (Japan), together with NRC (Canada) and NSC
and ASIAA (Taiwan), in cooperation with the Republic of Chile. The Joint ALMA
Observatory is operated by ESO, AUI/NRAO and NAOJ. A.I. and J.M.C. acknowledge
support from NSF awards AST-1109334 and AST-1140063. This publication makes use
of data products from the Two Micron All Sky Survey, which is a joint project
of the University of Massachusetts and the Infrared Processing and Analysis
Center/California Institute of Technology, funded by the National Aeronautics
and Space Administration and the National Science Foundation.

\clearpage

\begin{deluxetable}{lccc}
\tablecaption{Observed Sources\label{tbl:sample}}
\tablehead{
  \colhead{Source}  &
  \multicolumn{2}{c}{Phase Center (J2000)} &
  \colhead{UT Date Observed}
  \\
  \cline{2-3}
  \colhead{}  &
  \colhead{Right Ascension} &
  \colhead{Declination} &
  \colhead{}
}
\startdata
$[$PBB2002$]$ J155624.8$-$222555   & 15:56:24.774 & $-$22:25:55.26 & 2012-08-24\\
$[$PBB2002$]$ J155706.4$-$220606   & 15:57:06.419 & $-$22:06:06.10 & 2012-08-24\\
$[$PBB2002$]$ J155729.9$-$225843   & 15:57:29.862 & $-$22:58:43.85 & 2012-08-24\\
$[$PBB2002$]$ J155829.8$-$231007   & 15:58:29.813 & $-$23:10:07.72 & 2012-08-24\\
$[$PZ99$]$ J160357.6$-$203105      & 16:03:57.677 & $-$20:31:05.51 & 2012-12-16\\
$[$PZ99$]$ J160421.7$-$213028      & 16:04:21.655 & $-$21:30:28.40 & 2012-08-27\\
$[$PBB2002$]$ J160525.5$-$203539   & 16:05:25.564 & $-$20:35:39.71 & 2012-08-24\\
$[$PBB2002$]$ J160532.1$-$193315   & 16:05:32.152 & $-$19:33:15.99 & 2012-08-24\\
$[$PBB2002$]$ J160600.6$-$195711   & 16:06:00.616 & $-$19:57:11.46 & 2012-08-27\\
ScoPMS 31                          & 16:06:21.963 & $-$19:28:44.56 & 2012-12-16\\
$[$PBB2002$]$ J160622.8$-$201124   & 16:06:22.781 & $-$20:11:24.28 & 2012-08-27\\
$[$PBB2002$]$ J160643.8$-$190805   & 16:06:43.860 & $-$19:08:05.56 & 2012-12-16\\
$[$PBB2002$]$ J160702.1$-$201938   & 16:07:02.118 & $-$20:19:38.77 & 2012-08-24\\
$[$PBB2002$]$ J160823.2$-$193001   & 16:08:23.245 & $-$19:30:00.95 & 2012-12-16\\
$[$PBB2002$]$ J160827.5$-$194904   & 16:08:27.520 & $-$19:49:04.72 & 2012-08-27\\
$[$PBB2002$]$ J160900.0$-$190836   & 16:09:00.020 & $-$19:08:36.80 & 2012-08-27\\
$[$PBB2002$]$ J160900.7$-$190852   & 16:09:00.761 & $-$19:08:52.68 & 2012-12-16\\
$[$PBB2002$]$ J160959.4$-$180009   & 16:09:59.341 & $-$18:00:09.08 & 2012-08-24\\
$[$PZ99$]$ J161411.0$-$230536      & 16:14:11.077 & $-$23:05:36.24 & 2012-08-27\\
$[$PBB2002$]$ J161420.2$-$190648   & 16:14:20.299 & $-$19:06:48.14 & 2012-08-27
\enddata
\end{deluxetable}


\clearpage
\begin{deluxetable}{lrcccr}
\tablecaption{Stellar Properties\label{tbl:stellar}}
\tablehead{
  \colhead{Source}  &
  \colhead{SpT}     &
  \colhead{$\mathrm{A_v}$} &
  \colhead{log ($T_*$/K)}  &
  \colhead{log ($L_*$/L$_\odot$)}  &
  \colhead{log ($M_*$/M$_\odot$)}
}
\startdata
$[$PBB2002$]$ J155624.8-222555 &   M4 & $0.69 \pm 0.36$ & $3.514 \pm 0.019$ & $-1.18 \pm 0.14$ & -0.66 (-0.14, +0.10) \\
$[$PBB2002$]$ J155706.4-220606 &   M4 & $0.70 \pm 0.50^{\rm{a}}$ & $3.514 \pm 0.019$ & $-1.44 \pm 0.14$ & -0.68 (-0.18, +0.09) \\
$[$PBB2002$]$ J155729.9-225843 &   M4 & $0.70 \pm 0.50^{\rm{a}}$ & $3.514 \pm 0.019$ & $-1.33 \pm 0.14$ & -0.69 (-0.14, +0.12) \\
$[$PBB2002$]$ J155829.8-231007 &   M3 & $1.07 \pm 0.40$ & $3.533 \pm 0.018$ & $-1.31 \pm 0.14$ & -0.58 (-0.10, +0.12) \\
$[$PZ99$]$ J160357.6-203105    &   K5 & $0.70 \pm 0.50^{\rm{a}}$ & $3.638 \pm 0.024$ & $-0.17 \pm 0.14$ &  0.01 (-0.05, +0.08) \\
$[$PZ99$]$ J160421.7-213028    &   K2 & $0.66 \pm 0.27$ & $3.690 \pm 0.016$ & $-0.24 \pm 0.14$ & -0.02 (-0.04, +0.05) \\
$[$PBB2002$]$ J160525.5-203539 &   M5 & $0.37 \pm 0.41$ & $3.495 \pm 0.020$ & $-1.37 \pm 0.14$ & -0.88 (-0.15, +0.17) \\
$[$PBB2002$]$ J160532.1-193315 &   M5 & $0.19 \pm 0.42$ & $3.495 \pm 0.020$ & $-1.59 \pm 0.14$ & -1.01 (-0.06, +0.25) \\
$[$PBB2002$]$ J160600.6-195711 &   M5 & $0.22 \pm 0.37$ & $3.495 \pm 0.020$ & $-1.20 \pm 0.14$ & -0.79 (-0.18, +0.11) \\
                     ScoPMS 31 & M0.5 & $0.98 \pm 0.26$ & $3.577 \pm 0.020$ & $-0.28 \pm 0.14$ & -0.31 (-0.09, +0.12) \\
$[$PBB2002$]$ J160622.8-201124 &   M5 & $0.00 \pm 0.37$ & $3.495 \pm 0.020$ & $-1.39 \pm 0.14$ & -0.89 (-0.17, +0.15) \\
$[$PBB2002$]$ J160643.8-190805 &   K6 & $0.72 \pm 0.25$ & $3.624 \pm 0.015$ & $-0.39 \pm 0.14$ & -0.04 (-0.05, +0.04) \\
$[$PBB2002$]$ J160702.1-201938 &   M5 & $0.92 \pm 0.37$ & $3.495 \pm 0.020$ & $-1.49 \pm 0.14$ & -0.90 (-0.17, +0.15) \\
$[$PBB2002$]$ J160823.2-193001 &   K9 & $1.05 \pm 0.29$ & $3.593 \pm 0.023$ & $-0.55 \pm 0.14$ & -0.13 (-0.16, +0.07) \\
$[$PBB2002$]$ J160827.5-194904 &   M5 & $0.70 \pm 0.39$ & $3.495 \pm 0.020$ & $-1.16 \pm 0.14$ & -0.78 (-0.18, +0.12) \\
$[$PBB2002$]$ J160900.0-190836 &   M5 & $0.42 \pm 0.35$ & $3.495 \pm 0.020$ & $-1.32 \pm 0.14$ & -0.84 (-0.20, +0.15) \\
$[$PBB2002$]$ J160900.7-190852 &   K9 & $1.32 \pm 0.25$ & $3.593 \pm 0.023$ & $-0.38 \pm 0.14$ & -0.08 (-0.23, +0.04) \\
$[$PBB2002$]$ J160959.4-180009 &   M4 & $0.56 \pm 0.36$ & $3.514 \pm 0.019$ & $-1.00 \pm 0.14$ & -0.59 (-0.17, +0.07) \\
$[$PZ99$]$ J161411.0-230536    &   K2 & $0.48 \pm 0.25$ & $3.690 \pm 0.030$ & $ 0.41 \pm 0.14$ &  0.20 (-0.07, +0.05) \\
$[$PBB2002$]$ J161420.2-190648 &   K5 & $2.00 \pm 0.50^{\rm{a}}$ & $3.638 \pm 0.024$ & $-0.29 \pm 0.14$ &  0.01 (-0.07, +0.04)
\enddata
\tablenotetext{a}{Photometry is not available to derive $\mathrm{A_v}$; the assumed value is listed (see text).}
\end{deluxetable}

\clearpage



\begin{deluxetable}{cccccccc}
\tablecaption{ALMA Observations\label{tbl:obs}}
\tablehead{
  \colhead{UT Date}  &
  \colhead{Number}     &
  \colhead{Baseline range} &
  \colhead{pwv} &
  \multicolumn{4}{c}{Calibrators}\\
  \cline{5-8}
  \colhead{}  &
  \colhead{Antennas}     &
  \colhead{(m)} &
  \colhead{(mm)} &
  \colhead{Flux} &
  \colhead{Passband} &
  \colhead{Secondary} &
  \colhead{Gain}
}
\startdata
2012 Aug 24 & 25 & 17--375 & 0.77 & Neptune & J1924$-$292 & J1751$-$0939 & J1625$-$2527\\
2012 Aug 28 & 23 & 12--386 & 0.68 & Titan   & J1924$-$292 & J1751$-$0939 & J1625$-$2527\\
2012 Dec 16 & 17 & 16--402 & 1.16 & Titan   & J1924$-$292 & \ldots       & J1625$-$2527
\enddata
\end{deluxetable}


\clearpage
\begin{deluxetable}{lcccccc}
\footnotesize
\tablecaption{Measured Continuum Flux Densities at a Mean Frequency of 340.7~GHz\label{tbl:continuum}}
\tablehead{
  \colhead{Source}  &
  \colhead{$S_\mathrm{total}$}   &
  \colhead{$\Delta\alpha$}     &
  \colhead{$\Delta\delta$}     &
  \colhead{$\sigma$} &
  \colhead{$\theta_\mathrm{b}$} &
  \colhead{P.A.}\\
  \colhead{}  &
  \colhead{(mJy)}     &
  \colhead{(arcsec)} &
  \colhead{(arcsec)} &
  \colhead{(mJy beam$^{-1}$)} &
  \colhead{(arcsec)} &
  \colhead{(deg)} \\
  \colhead{(1)} &
  \colhead{(2)} &
  \colhead{(3)} &
  \colhead{(4)} &
  \colhead{(5)} &
  \colhead{(6)} &
  \colhead{(7)}
}
\startdata
$\mathrm{[PBB2002]UScoJ155624.8-222555}$  &   0.28 $\pm$   0.18  & \ldots & \ldots  &   0.15  & $0.80\times0.48$  & $-74$ \\
$\mathrm{[PBB2002]UScoJ155706.4-220606}$  &   0.32 $\pm$   0.20  & \ldots & \ldots  &   0.20  & $0.91\times0.48$  & $-73$ \\
$\mathrm{[PBB2002]UScoJ155729.9-225843}$  &  -0.04 $\pm$   0.20  & \ldots & \ldots  &   0.19  & $0.86\times0.48$  & $-74$ \\
$\mathrm{[PBB2002]UScoJ155829.8-231007}$  &   5.86 $\pm$   0.18  & $0.10 \pm 0.11$ & $-0.01 \pm 0.11$  &   0.20  & $0.76\times0.48$  & $-75$ \\
$\mathrm{[PZ99]J160357.6-203105}$  &   4.30 $\pm$   0.39  & $0.01 \pm 0.08$ & $0.06 \pm 0.08$  &   0.45  & $0.65\times0.48$  & $-8$ \\
$\mathrm{[PZ99]J160421.7-213028}$\tablenotemark{a}  & 218.76 $\pm$   0.81  & $0.01 \pm 0.11$ & $-0.03 \pm 0.11$  &   0.19  & $0.76\times0.45$  & $-74$ \\
$\mathrm{[PBB2002]UScoJ160525.5-203539}$  &   1.53 $\pm$   0.20  & $-0.09 \pm 0.19$ & $0.52 \pm 0.19$  &   0.19  & $0.99\times0.48$  & $-72$ \\
$\mathrm{[PBB2002]UScoJ160532.1-193315}$  &   0.25 $\pm$   0.20  & \ldots & \ldots  &   0.23  & $0.94\times0.48$  & $-72$ \\
$\mathrm{[PBB2002]UScoJ160600.6-195711}$  &  -0.00 $\pm$   0.13  & \ldots & \ldots  &   0.13  & $0.68\times0.45$  & $-76$ \\
$\mathrm{ScoPMS31}$             &   4.08 $\pm$   0.52  & $0.02 \pm 0.22$ & $0.50 \pm 0.22$  &   0.53  & $0.65\times0.49$  & $-3$ \\
$\mathrm{[PBB2002]UScoJ160622.8-201124}$  &   0.59 $\pm$   0.14  & $0.09 \pm 0.19$ & $0.05 \pm 0.19$  &   0.13  & $0.70\times0.45$  & $-75$ \\
$\mathrm{[PBB2002]UScoJ160643.8-190805}$  &   1.11 $\pm$   0.42  & \ldots & \ldots  &   0.47  & $0.65\times0.48$  & $-5$ \\
$\mathrm{[PBB2002]UScoJ160702.1-201938}$  &  -0.09 $\pm$   0.20  & \ldots & \ldots  &   0.20  & $1.04\times0.48$  & $-72$ \\
$\mathrm{[PBB2002]UScoJ160823.2-193001}$\tablenotemark{b}  &  43.19 $\pm$   0.81  & $0.21 \pm 0.20$ & $0.29 \pm 0.21$  &   0.52  & $0.65\times0.50$  & $-9$ \\
$\mathrm{[PBB2002]UScoJ160827.5-194904}$  &   0.76 $\pm$   0.13  & $0.01 \pm 0.15$ & $-0.03 \pm 0.15$  &   0.16  & $0.64\times0.45$  & $-76$ \\
$\mathrm{[PBB2002]UScoJ160900.0-190836}$  &   1.73 $\pm$   0.13  & $0.04 \pm 0.12$ & $0.09 \pm 0.12$  &   0.14  & $0.66\times0.45$  & $-76$ \\
$\mathrm{[PBB2002]UScoJ160900.7-190852}$\tablenotemark{b}  &  47.28 $\pm$   0.91  & $0.42 \pm 0.20$ & $-0.27 \pm 0.21$  &   0.62  & $0.65\times0.48$  & $-4$ \\
$\mathrm{[PBB2002]UScoJ160959.4-180009}$  &   0.67 $\pm$   0.18  & $-0.19 \pm 0.26$ & $-0.13 \pm 0.26$  &   0.18  & $0.80\times0.48$  & $-72$ \\
$\mathrm{[PZ99]J161411.0-230536}$  &   4.77 $\pm$   0.14  & $0.09 \pm 0.04$ & $-0.07 \pm 0.04$  &   0.16  & $0.70\times0.45$  & $-76$ \\
$\mathrm{[PBB2002]UScoJ161420.2-190648}$\tablenotemark{b}  &  40.69 $\pm$   0.22  & $-0.12 \pm 0.20$ & $0.11 \pm 0.20$  &   0.15  & $0.77\times0.45$  & $-73$
\enddata
\tablenotetext{a}{Integrated flux density measured on an image with a circular aperture of 1.5\arcsec\ radius.}
\tablenotetext{b}{Integrated flux density measured by fitting an elliptical gaussian to the visibility data.}
\tablecomments{\\
Column (1) : star name;\\
Column (2) : integrated flux density derived by fitting a point source model to the $uv$ data, unless otherwise indicated;\\
Column (3) and (4) : right ascension and declination offsets of the ALMA continuum source from the stellar position; ellipses indicate that the source was not detected with ALMA and the offsets wered fixed at the stellar position during the model fitting;\\
Column (5) : RMS noise in an image created with robust=2 and measured in an annulus between 4\arcsec\ and 5\arcsec\ centered on the stellar position;\\
Column (6) : FWHM synthesized beam size;\\
Column (7) : position angle of the beam measured east of north.\\
}
\end{deluxetable}

\clearpage
\begin{deluxetable}{lc}
\tablecaption{Derived Dust Masses\label{tbl:mdust}}
\tablehead{
  \colhead{Source}  &
  \colhead{$\mathrm{log (M_{dust} / M_\oplus})$}
}
\startdata
$\mathrm{[PBB2002]UScoJ155624.8-222555}$ & $< -0.34$ \\
$\mathrm{[PBB2002]UScoJ155706.4-220606}$ & $< -0.17$ \\
$\mathrm{[PBB2002]UScoJ155729.9-225843}$ & $< -0.40$ \\
$\mathrm{[PBB2002]UScoJ155829.8-231007}$ & $0.58 \pm  0.13$ \\
$\mathrm{[PZ99]J160357.6-203105}$ & $-0.01 \pm  0.14$ \\
$\mathrm{[PZ99]J160421.7-213028}$ & $1.72 \pm  0.13$ \\
$\mathrm{[PBB2002]UScoJ160525.5-203539}$ & $0.02 \pm  0.14$ \\
$\mathrm{[PBB2002]UScoJ160532.1-193315}$ & $< -0.12$ \\
$\mathrm{[PBB2002]UScoJ160600.6-195711}$ & $< -0.65$ \\
$\mathrm{ScoPMS31}$ & $0.00 \pm  0.14$ \\
$\mathrm{[PBB2002]UScoJ160622.8-201124}$ & $-0.38 \pm  0.17$ \\
$\mathrm{[PBB2002]UScoJ160643.8-190805}$ & $< -0.19$ \\
$\mathrm{[PBB2002]UScoJ160702.1-201938}$ & $< -0.33$ \\
$\mathrm{[PBB2002]UScoJ160823.2-193001}$ & $1.13 \pm  0.13$ \\
$\mathrm{[PBB2002]UScoJ160827.5-194904}$ & $-0.38 \pm  0.15$ \\
$\mathrm{[PBB2002]UScoJ160900.0-190836}$ & $0.05 \pm  0.13$ \\
$\mathrm{[PBB2002]UScoJ160900.7-190852}$ & $1.10 \pm  0.13$ \\
$\mathrm{[PBB2002]UScoJ160959.4-180009}$ & $-0.50 \pm  0.17$ \\
$\mathrm{[PZ99]J161411.0-230536}$ & $-0.16 \pm  0.13$ \\
$\mathrm{[PBB2002]UScoJ161420.2-190648}$ & $1.01 \pm  0.13$
\enddata
\end{deluxetable}

\end{document}